\documentclass[twocolumn, trackchanges]{aastex63}

\submitjournal{ApJ}

\usepackage{subfiles}
\usepackage{CJK}
\usepackage{amsmath}
\usepackage{enumitem}

\setlist[enumerate]{nosep}
\usepackage{cleveref}

\newcommand{\athenapp}{\texttt{Athena++}~}

\begin{document}
\begin{CJK*}{UTF8}{gbsn}

\title{Local Simulations of Heating Torques on a Luminous Body in an Accretion Disk}

\correspondingauthor{Amelia (Lia) Hankla}
\email{lia.hankla@gmail.com, amelia.hankla@colorado.edu}
\author[0000-0001-9725-5509]{Amelia M. Hankla}
\altaffiliation{National Science Foundation Graduate Research Fellow}
\affiliation{JILA, University of Colorado and National Institute of Standards and Technology, 440 UCB, Boulder, CO 80309-0440, USA}
\affiliation{Department of Physics, University of Colorado, 390 UCB, Boulder, CO 80309-0390, USA}

\author[0000-0002-2624-3399]{Yan-Fei Jiang(姜燕飞)}
\affiliation{Center for Computational Astrophysics, Flatiron Institute, NY 10010, USA}

\author[0000-0001-5032-1396]{Philip J. Armitage}
\affiliation{Center for Computational Astrophysics, Flatiron Institute, NY 10010, USA}
\affiliation{Department of Physics and Astronomy, Stony Brook University, Stony Brook, NY 11794, USA}

\begin{abstract}
A luminous body embedded in an accretion disk can generate asymmetric density perturbations that lead to a net torque and thus orbital migration of the body. Linear theory has shown that this heating torque gives rise to a migration term linear in the body's mass that can oppose or even reverse that arising from the sum of gravitational Lindblad and co-orbital torques. We use high-resolution local 3D shearing sheet simulations of a zero-mass test particle in an unstratified disk to assess the accuracy and domain of applicability of the linear theory. We find agreement between analytic and simulation results to better than 10\% in the low luminosity, low thermal conductivity regime, but measure deviations in both the non-linear (high luminosity) and the high thermal conductivity regimes. In the non-linear regime, linear theory overpredicts the acceleration due to the heating torque, potentially due to the neglect of non-linear terms in the heat flux. In the high thermal conductivity regime linear theory underpredicts the acceleration, which scales with a power-law index of $-1$ rather than $-3/2$, although here both non-linear and computational constraints play a role. We discuss the impact of the heating torque for the evolution of low-mass planets in protoplanetary disks, and for massive stars or accreting compact objects embedded in Active Galactic Nuclei disks. For the latter case, we show that the thermal torque is likely to be the dominant physical effect at disk radii where the optical depth drops below $\tau\lesssim 0.07\alpha^{-3/2}\epsilon c/v_K$.
\end{abstract}

\section{Introduction}
Planets, stars or compact objects orbiting within accretion disks perturb surrounding gas due to gravitational forces \citep{goldreich80}, accretion \citep{bondi52}, the release of heat or radiation \citep{Lega14,BL15, masset17}, and winds \citep{gruzinov19}. It is commonly the case that the resulting density perturbations leading and trailing the orbital motion are asymmetric, producing a gravitational back-reaction and a non-zero torque on the body. The torque leads to orbital migration---either an increase or a decrease in the semi-major axis---and evolution of any eccentricity or inclination (usually in the sense of damping, but in the opposite sense when combined with the release of heat \citep{Eklund2017}). In many circumstances of interest the time scale for migration is short compared to the disk lifetime, making it probable that observable properties of the system are substantially shaped by the effects of migration. 

The longest-studied torque is that due to the purely gravitational perturbation of the disk-embedded object. It is made up of two independent components, one from waves excited at Lindblad resonances and one exerted in the co-orbital region \citep{kley12}, both of which scale as the square of the object's mass. The net Lindblad torque (summing the opposite-signed contributions from interior and exterior resonances) has some dependence on disk properties, but is mostly due to intrinsic asymmetries in the interaction and almost always leads to inward migration \citep{ward97}. The co-orbital torque, on the other hand, can lead to either inward or outward migration, and depends in a complex way on numerous properties of the disk \citep[including radial gradients of vortensity and entropy, viscosity, thermal diffusivity, and disk winds;][]{paardekooper11,mcnally20}.

Numerical simulations show that thermal effects, either in the disk gas in the vicinity of the planet or associated with the release of heat or radiation from a luminous body, result in additional torques \citep{Lega14, BL15,chrenko17}. Unlike the purely gravitational torques, thermal effects can (in principle) remain significant even for very low mass bodies. In particular, \citet{masset17}, using linear perturbation theory, identified a ``heating torque" that arises when an orbiting body injects thermal energy into the surrounding disk. The thermal energy leads to the formation of low-density lobes near the planet, which are generically asymmetric, producing a torque. This heating torque is due purely to the injection of luminosity into the surrounding gas; the effect of the planet's gravitational potential leads to another type of thermal torque called the ``cold thermal torque". The cold and the heating thermal torques can be separated and studied independently in the linear regime, that is, whereas the cold thermal torque is only present for a massive planet, the heating torque can be studied for a massless planet. Both the heating torque and the cold thermal torque can be on the same order of magnitude as other torques that cause migration (such as the Lindblad torque), and typically lead to outward migration. 

The consequences of thermal torques on the migration rate of disk-embedded objects have been studied in the context of low-mass planet formation, where Lindblad torques alone would cause planets with masses of the order of the Earth's mass to migrate toward the central star on a timescale shorter than the disk lifetime. The luminosity on these mass scales typically results from pebble accretion \citep{ormel10,lambrechts12}. The heating torque modifies the predicted map of where in the disk inward and outward migration occur \citep{guilera19}, though the consequences for the final population of planets that form may be modest \citep{baumann20}.

Heating torques could also impact the migration rate of luminous bodies such as stars and accreting stellar-mass black holes, which can be captured \citep{syer91} or form \citep{shlosman87,goodman2003,levin07,dittmann20}
in the gas disks around supermassive black holes. Heating torques could interact with other gas torques (e.g. Lindblad torque, corotation torques) to form a migration trap---a radius in Active Galactic Nuclei (AGN) disk where the net torque is zero. Such migration traps would host an increased density of objects and provide a possible formation location for intermediate-mass black holes~\citep{bellovary16} or for stellar-mass black hole  binaries~\citep{secunda19,tagawa19}. Stellar-mass binaries merging {\em within} an AGN disk could contribute to the observed LIGO population \citep{stone17,abbott19}, while stellar-mass black holes merging with the central supermassive black hole are future LISA sources, whose detailed properties may be modified by migration torques~\citep{derdzinski2019}. 
 
Heating torques have been studied analytically \citep{masset17} and using global numerical simulations \citep{Lega14,BL15,chrenko17}. Here, we complement these prior studies using a local shearing box model for the disk. By simulating a luminous body in the limit where its mass goes to zero, using 32 zones per characteristic wavelength of the heating torque, we are able (a) to isolate the heating torque from the cold thermal torque and (b) to fully resolve the influence of the heating torque on the disk. Our work effectively extends the thorough numerical investigation of a luminous body travelling through a homogeneous medium~\citep{velascoromero19,velascoromero20} to the case of a luminous body embedded within a shear flow. An important difference between these past studies and the present study is the massless nature of our planet, a piece of physics which we do not consider because the so-called ``cold thermal torque" due to the gravitational potential of the planet should be separable from the heating torque in the linear regime~\citep{masset17}. The main questions we seek to answer are:
\begin{enumerate}
    \item What are the numerical prerequisites needed to reproduce the~\citep{masset17} linear theory, and how accurate is that theory when the approximations involved are relaxed?
    \item When do non-linear effects set in, and how do they change the linear theory's prediction for the thermal torque?
    \item Is the heating torque important for stars and accreting compact objects embedded within AGN disks?
\end{enumerate}

The structure of the paper is as follows: we summarize the analytic 
theory for the heating torque resulting from a luminous body in a shear flow in \S\ref{ssec:analyticsummary} and describe our numerical methods in \S\ref{ssec:simulations}. Our numerical results are presented in \S\ref{sec:results}.  \S\ref{ssec:limitations} discusses the limits to the analytic theory, and \S\ref{ssec:scenarios} discusses applications of the model to luminous objects in AGN and protoplanetary disks. We conclude in \S\ref{sec:conclusions}.

\section{Methods}
\label{sec:methods}
\subsection{Analytic Results} 
\label{ssec:analyticsummary}
\begin{figure*}[ht!]
    \centering
    \includegraphics[width=0.8\textwidth]{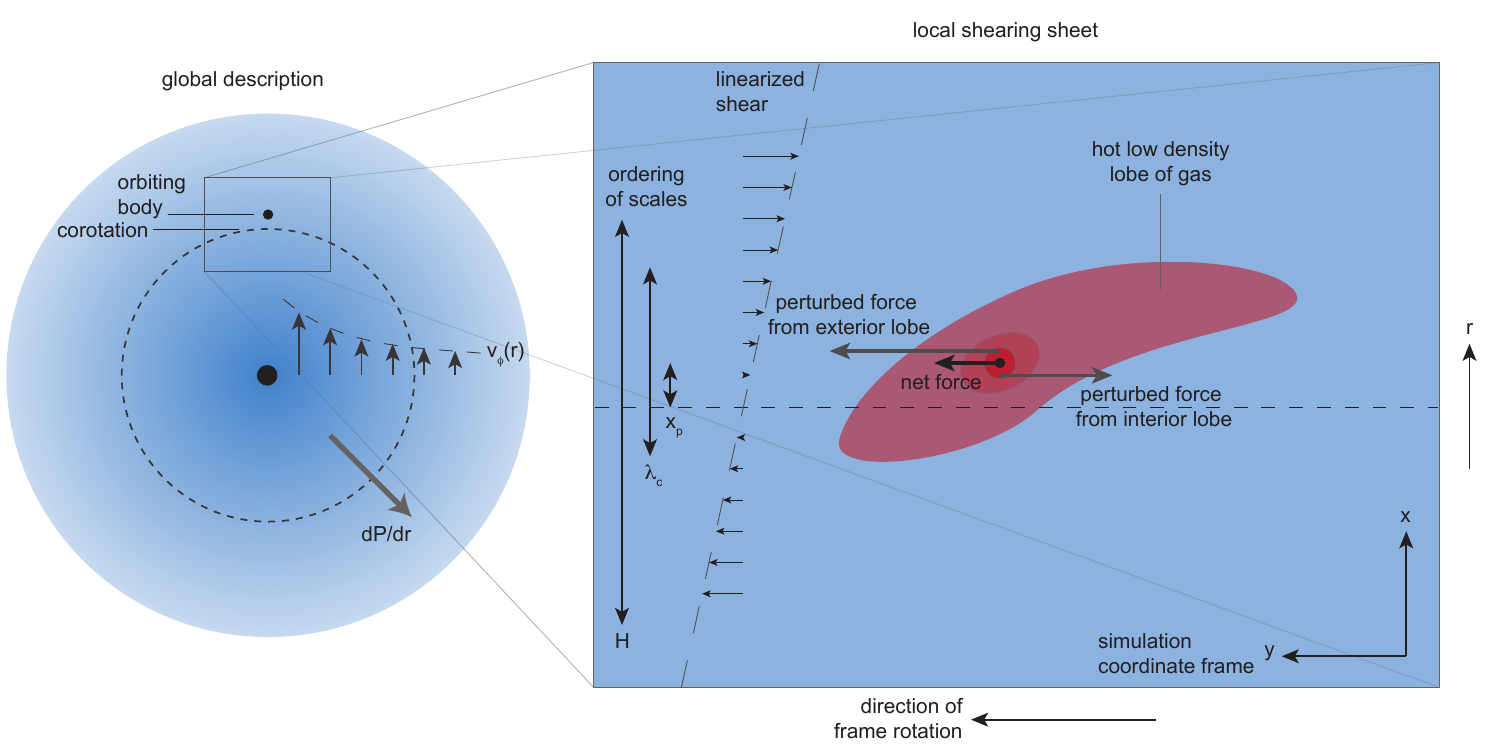}
    \caption{Illustration of the physics leading to a heating torque (the gravitational potential of the body is neglected). Heat diffusing away from a luminous, disk-embedded body, is sheared by the sub-Keplerian disk flow, forming hot low density lobes. These lobes are asymmetric interior to / exterior to the body, because the body is displaced from exact co-rotation due to the presence of a pressure gradient in the disk gas. The gravitational back-reaction from the heated lobes exerts a positive torque on the orbiting body, as illustrated by the vectors showing the \textit{perturbed} force in the azimuthal direction, i.e. the difference between the force on a luminous massless body and the force on a non-luminous massless body. Other components of the force (e.g. in the radial direction) are not drawn because they are opposite and equal and thus do not contribute to the total torque}.
    \label{fig:cartoon}
\end{figure*}

Analysis of the local hydrodynamic equations with thermal conductivity shows that the effects of a massive body's gravitational potential and its luminosity on a surrounding disk can be separated and studied independently in the linear regime~\cite[Eq. 34]{masset17}. We take advantage of this separation to focus solely on the ``heating torque", the torque due to the density perturbation that is sourced by thermal energy diffusing outward from a luminous body, though in the presence of orbital eccentricities and inclinations the applicability of the linear regime is limited \citep{Eklund2017,chrenko17,fromenteau2019}. Fig.~\ref{fig:cartoon} illustrates how the asymmetry in this perturbation, resulting from the displacement of the orbiting body from co-rotation, leads to a net torque. To aid in the interpretation of our numerical results, we summarize the key assumptions and results from \citet{masset17}.

\citet{masset17} linearizes the hydrodynamic equations (see~\crefrange{eq:masseq}{eq:energyeq}), assuming a steady-state, in a local (``shearing box") frame that co-rotates with the orbiting body. In this local frame, $x$ corresponds to the radial direction, $y$ to the azimuthal, and $z$ to the vertical direction (perpendicular to the disk midplane) as illustrated in Fig.~\ref{fig:cartoon}. If there is a radial pressure gradient in the disk, there is an offset $x_p$ between the orbiting body and disk gas that has the same orbital velocity. This distance from co-rotation is given by,
\begin{equation}
   x_p=-\frac{\partial_x p_0}{2q\Omega_0^2\rho_0}. 
\label{eq:xp}
\end{equation}
Here $p_0$ ($\rho_0$) is the equilibrium background pressure (density), $q$ is the shearing parameter (equal to $3/2$ for Keplerian disks), and $\Omega_0$ is the angular velocity of the local frame. For typical pressure profiles that decrease as a function of radius $x_p$ is positive, implying that the body will sit further away from the central body than the gas rotating at the same angular velocity, experiencing a headwind. 

Three characteristic scales enter the problem: the distance from co-rotation $x_p$, the characteristic size of the density perturbation caused by the body's luminosity $\lambda_c$, and the pressure scale height of the disk $H$. In the linear calculation it is assumed that the following hierarchy holds,
\begin{align}
    x_p &\ll \lambda_c \label{eq:hierarchyI},  \\
    \lambda_c &\ll H \label{eq:hierarchyII}. 
\end{align}
We refer to the first requirement for scale separation (Eq.~\ref{eq:hierarchyI}) as Assumption II and the second (Eq.~\ref{eq:hierarchyII}) as Assumption III (the first assumption is that of linearity). Assumption III allows the vertical density gradient of the local box to be neglected (justifying our use of unstratified simulations, although stratification can cause oscillatory torques when the opacity is not constant;~\cite{Chrenko2019}), while the small parameter associated with Assumption II is used extensively to expand the expected gravitational force from the under-density caused by the body's luminosity. The relative importance of these two hierarchies and the validity of the predicted net force on the body (Eq.~\ref{eq:netfy}) is explored in \S~\ref{sssec:hierarchyTest}.

The characteristic size of the disturbance and the net azimuthal force experienced by the body as a result of the heating torque mechanism are predicted to be~\cite[Eq. 83, and Eq. 109]{masset17},
\begin{eqnarray}
    \lambda_c &=& 2\pi k_c^{-1}=2\pi\sqrt{\frac{\chi}{q\Omega_0\gamma}}, \label{eq:lambdac} \\
     F_y &=& \frac{0.322x_p\gamma^{3/2}(\gamma-1)GMLq^{1/2}\Omega_0^{1/2}}{\chi^{3/2}c_s^2}. \label{eq:netfy}   
\end{eqnarray}
Here $\gamma$ is the adiabatic index ($\gamma=5/3$ in all of the following simulations), $c_s^2=\gamma p_0/\rho_0$ is the equilibrium sound speed, $L$ is the luminosity emitted by the body, $\chi$ is the disk's thermal conductivity, and $M$ is the mass of the body. Crucially, Eq.~\ref{eq:netfy} is {\em linear} in the mass of the body. This feature of the heating torque allows us to calculate the force per unit mass (i.e. the body's acceleration) without needing to explicitly include the body's mass at all in the simulations.

The heating torque is of interest because it can be the same order of magnitude as the other torques in the system (such as the Lindblad torque). Defining
\begin{equation}
    L_c= \frac{4\pi GM\chi\rho_0}{\gamma},
\end{equation}
the heating torque can be written as~\citep[Eq.~144]{masset17}
\begin{equation}
    \Gamma^{heat}=1.61\frac{\gamma-1}\gamma\frac{x_p}{\lambda_c/2\pi}\frac L{L_c}\Gamma_0, \label{eq:htorquemag}
\end{equation}
where
\begin{equation}
    \Gamma_0=\sqrt{2\pi}\rho_0H r_0^4\Omega_0^2\left(\frac{M}{M_*}\right)^2\left(\frac{r_0}{H}\right)^3, \label{eq:lindblad}
\end{equation} 
is of the order of the Lindblad torque. Here $r_0$ is the semi-major axis of the body, $M_*$ is the mass of the central object, and $H$ is the pressure scale height of the disk. Note that this definition of $\Gamma_{0}$ differs from the more widespread definition (in for instance~\cite{paardekooper11}) by a factor of $r_{0}/H$.

In summary, the formula given by Eq.~\ref{eq:netfy} for the net force on a body (due to the asymmetric gravitational forces caused by the body's luminosity distributed by differential rotation) is predicted to hold under three conditions:
\begin{itemize}%[label=\textbf{Assumption \Roman*}, align=left]
    \item {\bf Assumption I}: perturbations of density and pressure should be much less than equilibrium values (linearity, $\rho'\ll\rho_0$). \label{a1}
    \item {\bf Assumption II}: the offset from corotation $x_p$ should be much less than the size of the disturbance $\lambda_c$ ( $x_p\ll\lambda_c$ ). \label{a2}
    \item {\bf Assumption III}: the disturbance should be much smaller than the pressure scale height of the disk ($\lambda_c\ll H$). \label{a3}
\end{itemize}
In this work, we test the validity of the linear theory when one or more of these assumptions is violated.

% -------------------------------------------------------------- %
\subsection{Simulations} \label{ssec:simulations}
The linear theory is developed in the local ``shearing sheet" approximation~\citep{masset17}, which translates directly into a well-studied numerical set-up. We solve the inviscid hydrodynamic equations in a local approximation of a Cartesian box rotating around a massive body (a star or black hole, for instance) with orbital frequency $\Omega_0$, and add a source term to the energy density equation to model the luminosity. With $\rho$ the mass density, $e$ the energy density, $P$ the pressure, and $\textbf{V}$ the velocity, the equations read,
\begin{align}
\frac{\partial\rho}{\partial t}+\nabla\cdot\left(\rho\textbf{V}\right)&=0, \label{eq:masseq}\\
\frac{\partial \rho\textbf{V}}{\partial t}+\nabla\cdot\left(\rho\textbf{VV}+P\textbf{I}\right) &= -\rho\nabla\Phi_t-2\rho\Omega~{\bf \hat z}\times\textbf{V}, \label{eq:momeq}\\
\frac{\partial e}{\partial t}+\nabla\cdot\left[\textbf{V}(e+P)+{\bf F_H}\right]&=L~\delta({\bf x}-{\bf x}_p),\label{eq:energyeq}
\end{align}
where ${\bf F_H}=-\chi\rho\nabla\left( e / \rho \right)$ is the heat flux and $L$ is the total luminosity emitted by the body and $\delta({\bf x}-{\bf x}_p)$ is the Dirac delta function. The gas has an adiabatic equation of state. $\Phi_t=-q\Omega_0^2(x-x_p)^2$ is the tidal potential due to the central object. The vertical density gradient is neglected, both for consistency with \citet{masset17} and for the same physical reasons discussed there, and the radial density gradient is modelled through a non-zero offset from co-rotation (it is neglected in the shearing-box, as justified by assuming that the background pressure does not change significantly over the short radial scales under consideration). The shearing parameter $q$ is equal to $3/2$ for the Keplerian flows studied in this work. 

The \athenapp code is used to solve the above equations in the luminous body's rest frame on a uniform Cartesian mesh~\citep[Stone et al., 2020, submitted]{white16, felker18} with the  Harten-Lax-van Leer-Contact Riemann solver. The simulation's origin sits at the radial location where the gas orbits at the same frequency as the luminous body, such that the position of the body is fixed over the course of the simulation (10 orbits). We discuss the consequences of neglecting the radial motion of the body in response to the generated torque in \S\ref{sssec:blcompare}. Both the origin and the body sit at the mid-plane of the disk ($z=0$). In the fiducial run L1K1, the simulation domain spans $[4.13,~12.4,~4.13]H$ in the $x$ (radial), $y$ (azimuthal), and $z$ (vertical) directions respectively, where $H$ is the pressure scale height of the disk (defined through $H=c_s/\Omega_0$). The fiducial run L1K1 has a resolution of $[256,~192,~256]$ cells (i.e. $[62.0,~20.7,~62.0]$ cells/$H$) and a value $x_p=0.097~H$. Convergence with resolution and domain size is studied in \S\ref{ssec:numerical}.

The body's luminosity is modelled by directly injecting internal energy density into the gas via the energy density equation (\ref{eq:energyeq}). In the analytic theory the injection term is $L~\delta ({\bf x} - {\bf x}_p)$ (as in Eq.~\ref{eq:energyeq}). Since injection at a single point is not possible numerically, we implement this term in the simulations by adding an energy density $\ell_v\times\Delta t$ at each time to each cell whose center lies within an injection radius $r_{\mathrm{rad}}$. Here, $\ell_v$ is the (constant) luminosity per volume and $\Delta t$ is the time step as determined by the Courant condition. The total luminosity $L$ injected at each time step can be calculated as $L=\ell_v\times n\times v$, where $n$ is the number of cells included in the injection region and $v$ is each cell's volume; thus the total luminosity $L$ depends on both the luminosity per volume $\ell_v$ and the injection radius $r_{\mathrm{rad}}$. Unless otherwise specified, the injection radius is set such that the luminosity is evenly distributed into the eight cells neighboring the body. The effect of a finite injection radius (rather than a strict Dirac delta function) is explored in \S\ref{ssec:numerical}. We note again that according to the linear theory the torque due to the gravitational potential of the body can be separated from the torque due to the body's luminosity~\citep{masset17}. In this work we model only the heating torque, and do not include the gravitational potential of the body. 

As is standard for the shearing-box set-up, all simulations use periodic boundary conditions in the azimuthal and vertical directions and shearing-periodic boundary conditions in the radial direction, the effects of which are discussed in \S\ref{ssec:numerical}. Generic units $H$ (disk scale height) for length, $\Omega^{-1}$ ($2\pi/\Omega$ is one orbit) for time, and $P_0$ (background pressure) for energy density are used. The sound speed thus has units of $H\Omega$, acceleration has units of $H\Omega^2$, thermal diffusivity has units of $H^2\Omega$, and total injected luminosity has units of $P_0H^3\Omega$. These values can be scaled to various astrophysical systems, which are discussed in \S\ref{sec:discussion}.

\subsection{Diagnostics} \label{ssec:diagnostics}
Linear theory provides a prediction for the total gravitational force $F_y$ experienced by the orbiting body as a result of the perturbed gas density (purely from the body's luminosity, not its gravity). We calculate this azimuthal force per unit body mass in the simulation on spherical shells by calculating the distance between the body and each cell, assuming that all of the cell's mass is located at its center, and using the y-component of the inverse square law with $GM=1$ in code-units. The resulting acceleration can be plotted as a function of radius or summed over radius to directly compare to Eq.~\ref{eq:netfy}. To avoid introducing artificial asymmetry to the force (i.e. a systematically larger force on the $x<0$ side), the summation stops at the shortest radius that fits inside the simulation domain in all directions. The limiting radius is thus $L_x - x_p$, where $L_x$ is the half-width of the box in the $x$ (radial) direction and $x_p$ is the body's distance from corotation. This restriction is not expected to change the results significantly since the excluded portions are not a large fraction of the box, the gas there is not as perturbed, and the force from the gas there is attenuated by the inverse square of the radius). 

The fractional change $\tilde f$ in a quantity $f$ is useful to establish the linearity of density perturbations,
\begin{equation}
    \tilde f=\frac{f(t)-f(t=0)}{f(t=0)}=\frac{f'(t)}{f(t=0)},
\end{equation}
where $f'(t)=f(t)-f(t=0)$ is the perturbation from equilibrium. We describe a simulation as being in the linear regime if the deviation from equilibrium values is no more than 5\%.

To facilitate direct comparison with previous work (specifically, \citet{masset17}'s Fig. 1), we calculate the perturbation in surface density $\sigma'$ as the $k_z=0$ mode of the Fourier transform $\hat \rho(x,y,k_z)$ of the density perturbation $\rho'$ in the $z$-direction, i.e.
\begin{align}
    \hat\rho(x,y,k_z)&=\int_{-\infty}^\infty\rho'(x,y,z)e^{-ik_zz}dz,\\
    \sigma(x,y)&=\hat\rho(x/k_c,y/k_c,0).
\end{align}
%\memopa{This feels like a really indirect way to say how we compute $\Sigma^\prime$. Couldn't we drop the exponential and just sum up $\rho^\prime$?}
Separating into the effect due to zero offset from corotation ($\sigma^{(0)})$ and the first order effect due to nonzero offset ($\sigma^{(1)}$), we find~\citep[Eq. 114]{masset17}
\begin{equation}
    \sigma'(x/k_c, y/k_c)=\sigma^{'(0)}(x/k_c,y/k_c)+x_p\sigma^{'(1)}(x/k_c,y/k_c).
\end{equation}
Note that $\sigma'$ is the \textit{perturbation} of the surface density; the unperturbed surface density $\sigma$ is constant. The expected offset of $\sigma^{'(1)}$ is $x_pk_c=0.59$ relative to~\citet{masset17}'s Fig. 1b. %\memopa{This construction is for comparison with Masset? I sort of see why we're doing this but it's rather confusing I think...}

% -------------------------------------------------------------- %
\section{Results} \label{sec:results}
\subsection{The Linear Regime} \label{ssec:linreg}
\begin{figure*}[ht!]
    \centering
    \includegraphics[width=\textwidth]{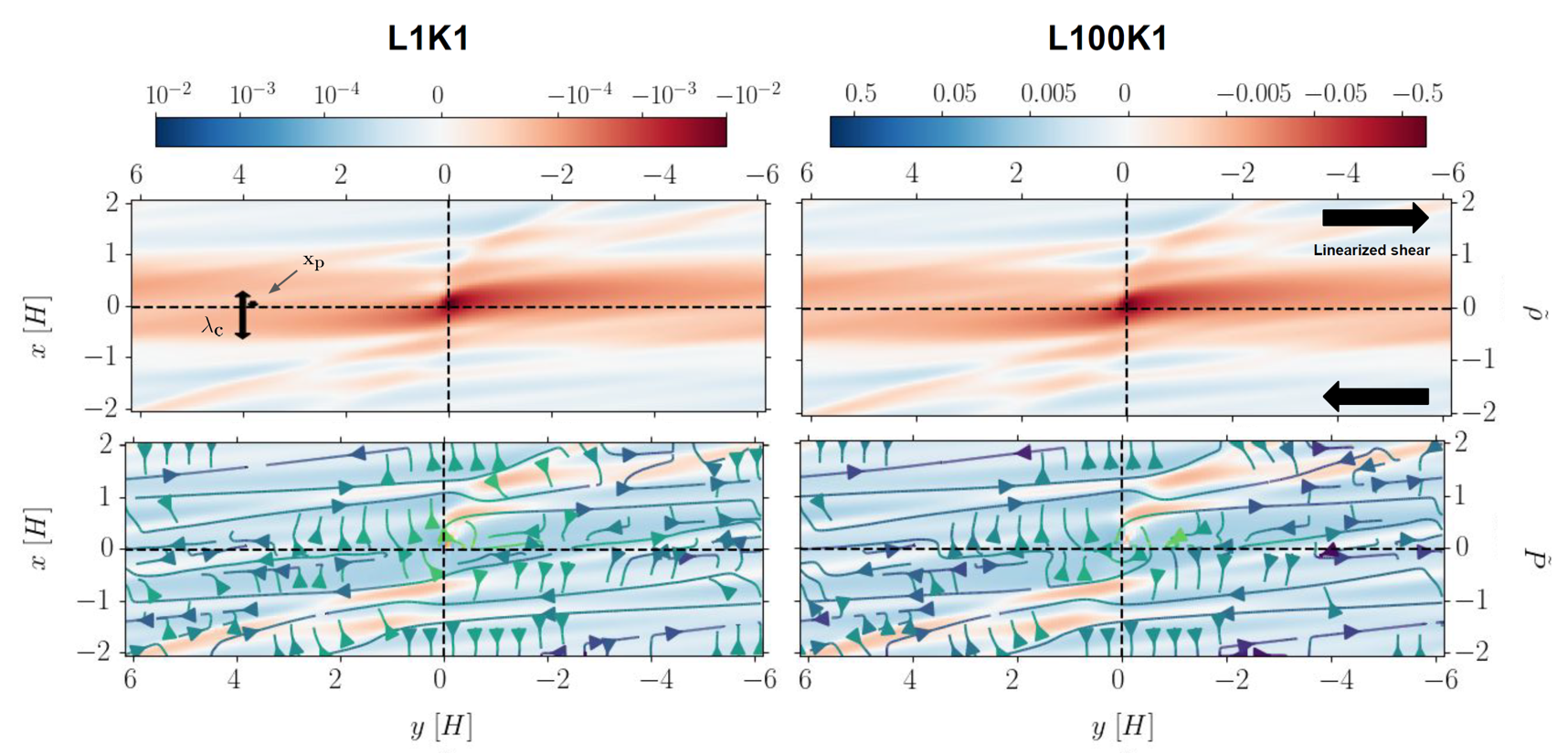}
    \caption{Slices in the $z=0$ plane of simulations with $\chi=0.017~H^2\Omega$ at $t=5.0$ orbits. Top row: density perturbation as a percentage of initial (equilibrium) condition. Bottom row: perturbation in pressure as a percentage of initial (equilibrium) condition. Left column: $\ell_v=1.42~P_0\Omega$ (fiducial simulation L1K1, linear regime). Right column: $\ell_v=142~P_0\Omega$ (high luminosity simulation L100K1, non-linear regime). Note that \bf{(though similar in shape)} the perturbations in L100K1 are larger in magnitude than the perturbations in L1K1. Velocity streamlines with the background shear profile subtracted are plotted in the lower panels, normalized to their maximum values for ease of comparison between L1K1 and L100K1. The distance from co-rotation $x_p$, the characteristic wavelength $\lambda_c$, and the direction of shear are marked with black arrows.}
    \label{fig:lin-nonlin}
\end{figure*}

Fig.~\ref{fig:lin-nonlin} shows the mid-plane pressure and density perturbations derived from simulations in the linear and non-linear regimes of luminosity injection. Values of $\ell_v=1.42~P_0\Omega$ (Fig.~\ref{fig:lin-nonlin} left column; physical values are discussed in ~\ref{sssec:blcompare}) lead to perturbations that are less than 5\% of the equilibrium values, which we take to be in the linear regime. The measured pressure perturbations are two orders of magnitude smaller than the density perturbations. This is consistent with \citet{masset17}'s estimation (Eq.~36) that $P'\ll H^2\Omega_0^2\rho'\approx c_s^2\rho'$. With the value $c_s^2=\gamma P/\rho = 1.00~H^2\Omega^2$, $P'$ should be much less than $\rho'$. The non-linear regime is illustrated in the right column of Fig.~\ref{fig:lin-nonlin}, which injects two orders of magnitude more energy per timestep ($\ell_v=142~P_0\Omega$; simulation L100K1). The qualitative appearance of the pressure and density perturbations remain similar for this much higher rate of energy injection. Both simulations, as expected, quickly reach an equilibrium within approximately two orbits.

\subsection{Net Azimuthal Acceleration as a Function of Radius}
\begin{figure}[ht!]
    \centering
    \includegraphics[width=0.5\textwidth]{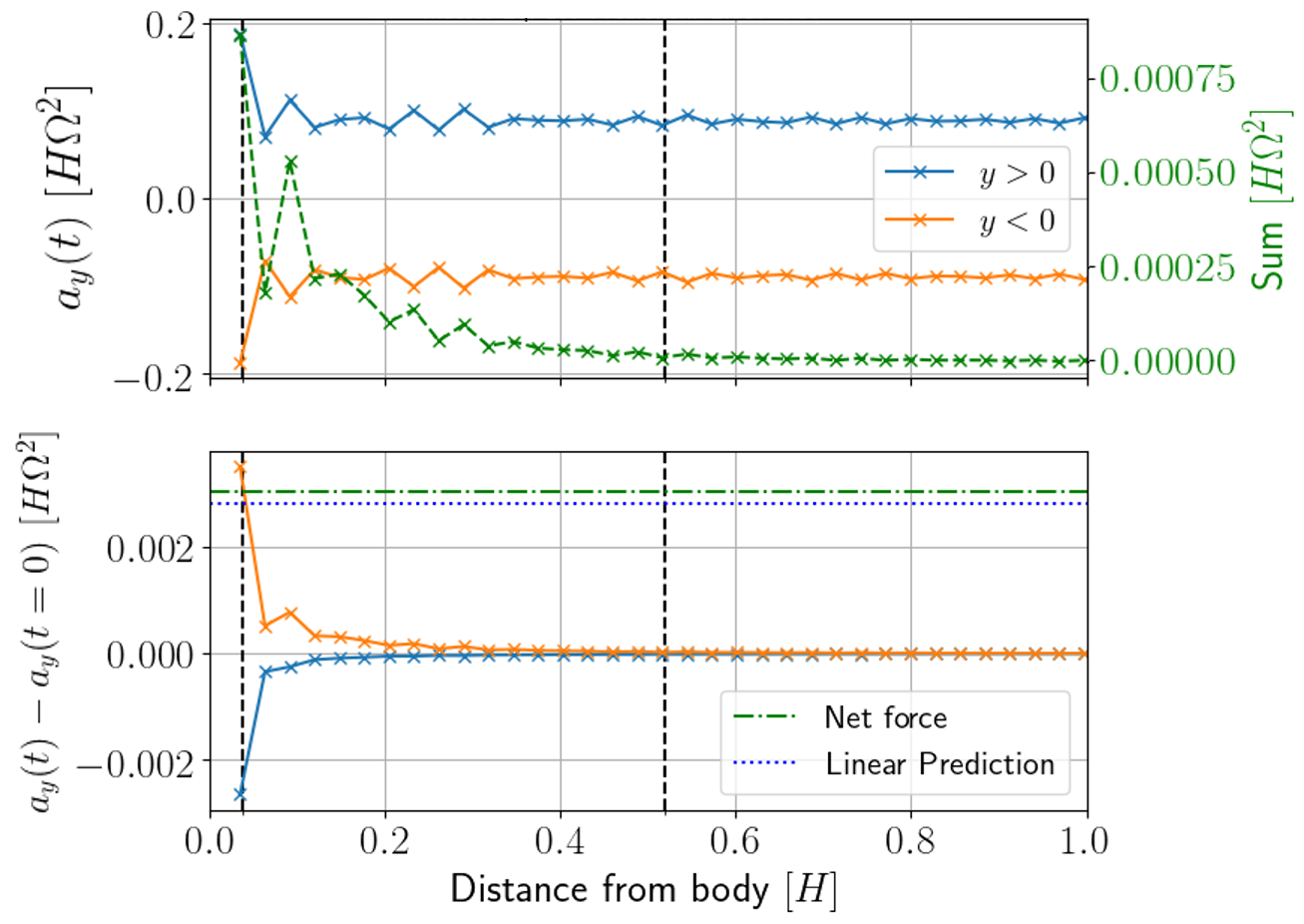}
    \caption{Snapshot of gravitational acceleration on the body in the fiducial simulation L1K1 as a function of distance away from the body at $t=5.0$ orbits. The top panel plots both the one-sided forces due to gas in front of ($y>0$) and behind ($y<0$) the body, as well as the sum of the forces (dashed green line; right scale). Vertical dashed lines show the size of the luminosity injection radius $r_\mathrm{rad}=0.04$ H and half the characteristic wavelength $\lambda_c=0.52~H$. The bottom panel shows the difference between the initial condition (which has net force equal to zero but one-sided forces on the order of the top panel's vertical values) and their values at five orbits. For reference the sum over all radii (with value $3.03\times10^{-3}~H\Omega^2$) is plotted as a dash-dot horizontal line, and for comparison the linear theory's predicted value of $2.82\times10^{-3}~H\Omega^2$ is plotted as a dotted horizontal line.}
    \label{fig:fy-rad}
\end{figure}

We take the $\ell_v = 1.42~P_0\Omega$ run as our fiducial simulation L1K1 so as to be firmly in the linear regime. Using the technique described in \S\ref{ssec:diagnostics}, we plot as a function of radius the net gravitational force on the body per unit body mass as a result of the gas perturbed by the object's luminosity. Fig.~\ref{fig:fy-rad} shows the result of summing up gas in front of and behind the body ($y>0$ and $y<0$, respectively), as well as the total force (green; right-hand scale). The bottom panel shows how the net azimuthal acceleration differs ahead/behind the body from the initial acceleration (which is non-zero on either side but which sums to zero) as well as how close the net azimuthal acceleration summed over all radii is to the linear prediction (horizontal lines). The noisiness of the data in Fig.~\ref{fig:fy-rad} is due to the fact that no interpolation was used in the calculation of the acceleration. Measurements from this simulation agree with the linear prediction to better than 10\%.

\subsection{Perturbation of the Surface Density}
\begin{figure}
    \centering
    \includegraphics[width=0.5\textwidth]{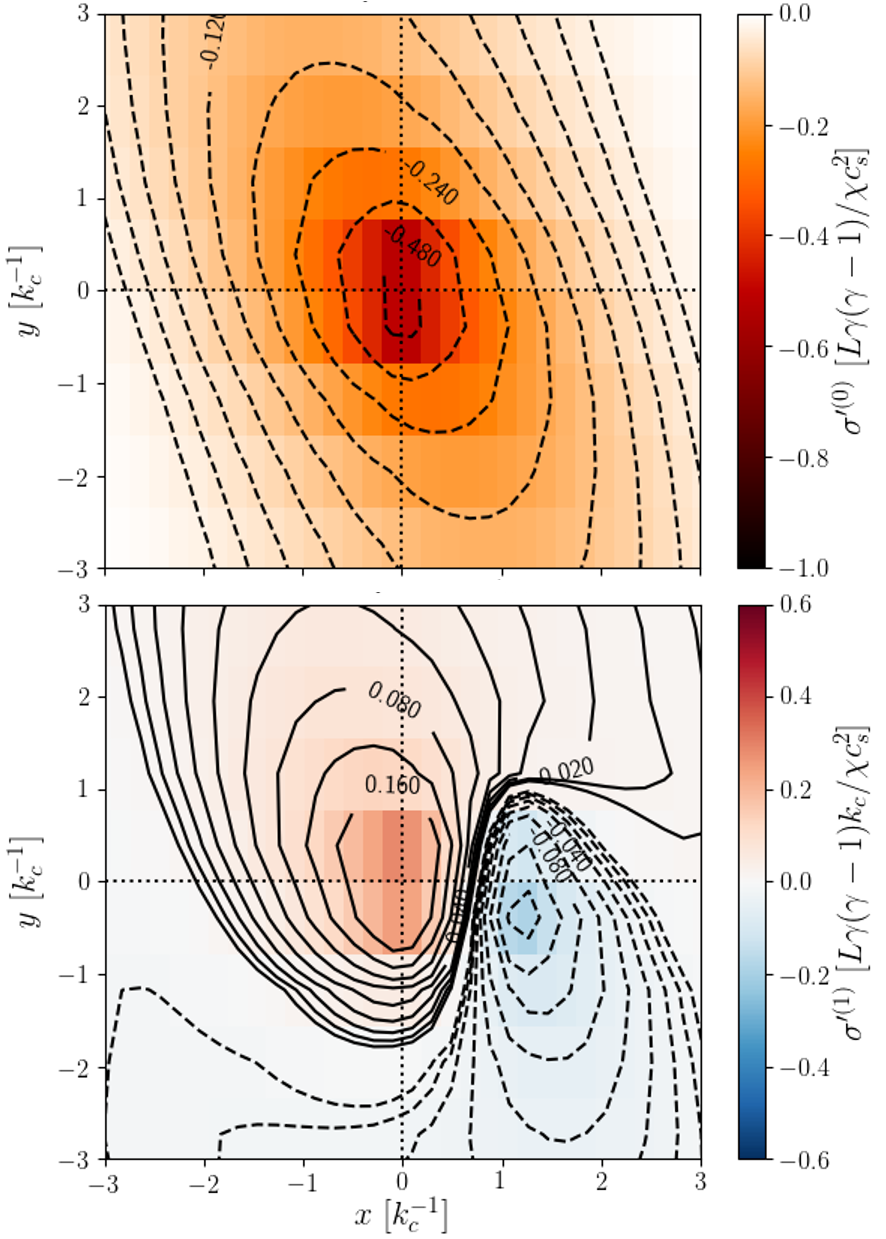}
    %\begin{subfigure}{0.5\textwidth}
    %\centering
    %\includegraphics[width=\textwidth]{masset-fig1a_t040.png}
    %\end{subfigure}%
    %\begin{subfigure}{0.5\textwidth}
    %\centering
    %\includegraphics[width=\textwidth]{masset-fig1b_t040.png}%
    %\end{subfigure}
    \caption{Perturbation of surface density in units of $\gamma(\gamma-1)L/\chi c_s^2$ due to the luminous body's heat  at $t=3.5$ orbits. Contour levels on the left are a geometric series with a ratio of $\sqrt{2}$ from $-0.03$ to $-0.48$. On the right, contour levels have a ratio of $2$ between them and run from $\pm0.01$ to $\pm0.16$. Solid contour are positive values; dashed are negative. Thermal conductivity is $\chi=0.017~H^2\Omega$, $\ell_v=1.42 P_0\Omega$. Similar to~\citet[Fig. 1]{masset17}.
    \label{fig:mfig1}
%{\color{red} YFJ"  May be better to use the same symbol for surface density as defined in equation 15, instead of $\sigma$.}    
    }
\end{figure}

In addition to the value of the net force acting on the body, \citet{masset17} calculates a map of the surface density perturbation predicted by the analytic theory. Fig.~\ref{fig:mfig1} reproduces this map with simulation data (compare to~\citet{masset17}'s Fig. 1). The upper panel Fourier transforms the density perturbation of a simulation with no offset ($x_p=0$), as outlined in \S\ref{ssec:diagnostics}. The lower panel Fourier transforms the density perturbation with the equilibrium value of $x_p$, then subtracts the zero offset case and divides by $x_p$ to extract $\sigma'^{(1)}$. The perturbations are smaller in amplitude than expected, largely because the peak expected amplitude is very close to the luminous body and is not as resolved. However, the general shape of the perturbation agrees well with the linear prediction.

\subsection{Scaling Relations} \label{ssec:scalingrel}
Linear theory predicts a linear dependence of the net gravitational force on the total luminosity $L$ emitted by the body and a power-law dependence $F_y\propto\chi^{-3/2}$ on the thermal conductivity (Eq.~\ref{eq:netfy}). To test these predictions we ran two suites of simulations: one that fixes the thermal conductivity and varies the total emitted luminosity, and one that fixes the total emitted luminosity and varies the thermal conductivity.

\begin{figure}
    \centering
    \includegraphics[width=0.48\textwidth]{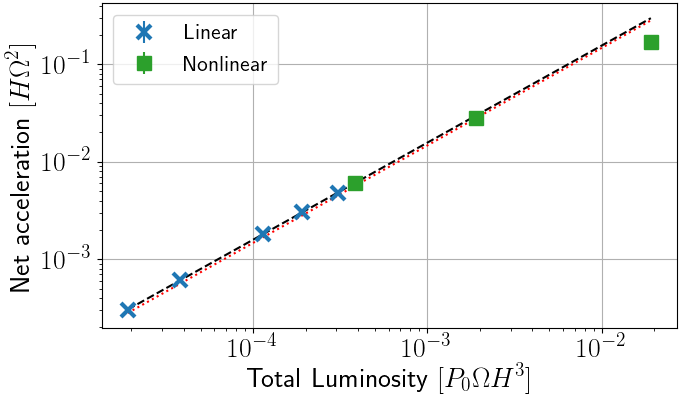}
    \caption{Net azimuthal acceleration due to the gas's gravity as a function of total injected luminosity. Thermal conductivity is fixed at $\chi=0.017~H^2\Omega$. The linear theory's prediction (slope: $1.0$; red dotted line) matches the data well even into the non-linear regime, although the fit (slope: $1.0\pm6.5\times10^{-7}$; black dashed line) was determined using only the linear data points. Simulations are summarized in Table~\ref{tab:fixedchisims}.}
    \label{fig:fixedchi}
\end{figure}

For the first suite we fixed $\chi=0.017~H^2\Omega$ and varied $\ell_v$ over three orders of magnitude: from $\ell_v=0.142~P_0\Omega$ to $142~P_0\Omega$. Simulations are considered to be in the linear regime if the perturbation never exceeds 5\% of the equilibrium value. Fig.~\ref{fig:fixedchi} reveals a tight agreement with the linear prediction even an order of magnitude into the non-linear regime (indicated by green squares). As the injected luminosity increases even more, the linear theory begins to over-predict the measured force because of non-linear effects; this over-prediction is discussed in \S\ref{sssec:non-linear}. In the linear regime at least, we are able to reproduce both the scaling and the normalization of the net force to within 10\%.

\begin{deluxetable*}{c|c|c|c|c} 
\tablecaption{Summary of simulations used to calculate the scaling of net azimuthal acceleration with total injected luminosity $L$ (Fig.~\ref{fig:fixedchi}). All simulations have $\chi=0.017~H^2\Omega$, corresponding to $\lambda_c=0.52~H$, while all other parameters (e.g. resolution, offset of the body from corotation, described in the text) are that of the fiducial simulation L1K1. This set of simulations keeps $\lambda_c/H$ and $x_p/\lambda_c$ constant at 0.52 and 0.187, respectively. * indicates a simulation that has density fluctuations greater than 5\% of the equilibrium value and has thus entered the non-linear regime. Simulation L1K1 is often referred to as the fiducial simulation, and L100K1 as the high luminosity simulation. Measured values are presented as the average between 1 and 10 orbits plus/minus one standard deviation of the value in time. \label{tab:fixedchisims}}
\tablewidth{0pt}
\tablehead{
\colhead{} & \colhead{$L~[P_0\Omega H^3]$} & \colhead{$\ell_v~[P_0\Omega]$} & \colhead{$F_y~[H\Omega^2]$ (Linear Prediction)} & \colhead{$F_y~[H\Omega^2]$ (Measured)}
}
  \startdata
    & 1.91$\times10^{-5}$ & 0.142 & 2.82$\times10^{-4}$ & 3.05$\times10^{-4}\pm$9.76$\times10^{-7}$ \\
    & 3.83$\times10^{-5}$ & 0.284 & 5.64$\times10^{-4}$ &  6.10$\times10^{-4}\pm$1.95$\times10^{-6}$ \\
    & 1.15$\times10^{-4}$ & 0.853 & 1.69$\times10^{-3}$ &  1.83$\times10^{-3}\pm$5.75$\times10^{-6}$ \\
    L1K1 & 1.91$\times10^{-4}$ & 1.42 & 2.82$\times10^{-3}$ &  3.03$\times10^{-3}\pm$9.46$\times10^{-6}$ \\ \hline
    & 3.06$\times10^{-4}$ & 2.28 & 4.51$\times10^{-3}$ &  4.83$\times10^{-3}\pm$1.50$\times10^{-5}$ \\
    * & 3.83$\times10^{-4}$ & 2.84 & 5.64$\times10^{-3}$ &  6.02$\times10^{-3}\pm$1.85$\times10^{-5}$ \\
    * & 1.91$\times10^{-3}$ & 14.2 & 2.82$\times10^{-2}$ &  2.83$\times10^{-2}\pm$8.02$\times10^{-5}$ \\
    * L100K1 & 1.91$\times10^{-2}$ & 142 & 2.82$\times10^{-1}$ &  1.69$\times10^{-1}\pm$4.59$\times10^{-4}$ \\
\enddata
\end{deluxetable*}

Assessing the validity of the analytic prediction for the scaling of the net force with thermal conductivity is substantially harder, because changing the conductivity also changes the characteristic wavelength $\lambda_c$. It is difficult to find a numerically tractable set of parameters that both (a) remains in the linear regime and (b) maintains the hierarchy of scales required by~\citet{masset17}, over a substantial range in $\chi$.

\begin{figure}
    \centering
    \includegraphics[width=0.48\textwidth]{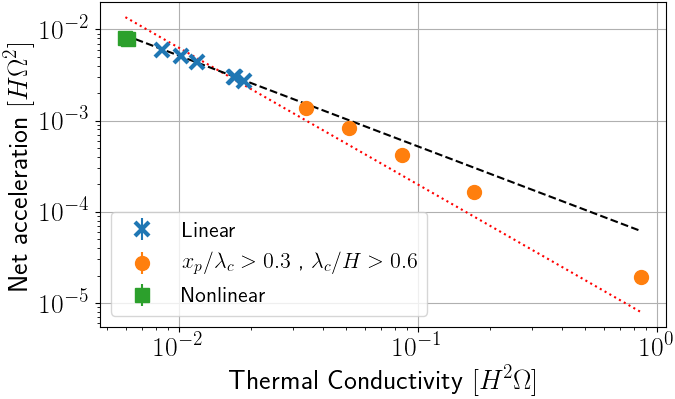}
    \caption{Net azimuthal acceleration due to the gas's gravity as a function of thermal conductivity. Total luminosity is fixed with $L=1.96\times10^{-4}~P_0\Omega H^3$ ($\ell_v=1.42~P_0\Omega$). The linear theory (red dotted line) predicts a power-law index of $-1.5$, whereas the fit (black dashed line) determines a power-law index of $-1.0\pm1.6\times10^{-4}$ and total luminosity $L=6.7\times10^{-3}~P_0\Omega H^3$. The fit was determined solely from the simulations satisfying the hierarchies $x_p/\lambda_c<0.3$ and $\lambda_c/H<0.6$ (blue crosses). Simulations are summarized in Table~\ref{tab:fixedLsims}.}
    \label{fig:fixedLtot}
\end{figure}

Fig.~\ref{fig:fixedLtot} shows the measured dependence of the net force as a function of the thermal conductivity, at fixed luminosity. For sufficiently low values of the thermal conductivity, heat cannot diffuse away fast enough, causing the system to enter the non-linear regime (indicated by a green square). For high values of the thermal conductivity, the required scale separation $x_p\ll \lambda_c\ll H$ is lost (shown as orange dots). Only the blue crosses, at intermediate $\chi$, remain linear and respect the scale hierarchy. 

Fitting the data only at intermediate $\chi$, we find that the dependence of net gravitational acceleration on conductivity is close to $\chi^{-1}$, rather than the expected $\chi^{-3/2}$. We caution, however, that this fit is made over only a very limited range of $\chi$. If, instead, we consider simulation data at higher values of $\chi$, we find a dependence that appears to be closer to the analytically predicted power-law. The ideal regime for matching the linear theory appears to be around $\chi=0.017~H^2\Omega$, which in the simulations presented has ratios $x_p/\lambda_c=0.187$ and $\lambda_c/H=0.519$. Simulations using $\chi=0.0061~H^2\Omega$ to obtain ratios $x_p/\lambda_c = 0.311 = \lambda_c/H$ measured an acceleration lower than the linear prediction by a factor of two. This suggests that the requirement that $x_p \ll\lambda_c$ is more important for matching linear theory than $\lambda_c\ll H$. This hierarchy of the assumption is reasonable since the former is used in expanding the net force, whereas the latter is used to drop vertical density stratification~\citep{masset17}; see \S\ref{sssec:hierarchyTest}.

\begin{deluxetable*}{c|c|c|c|c} 
\tablecaption{Summary of simulations used to calculate the scaling of net azimuthal acceleration with conductivity $\chi$ (Fig.~\ref{fig:fixedLtot}). All simulations have $L=1.9\times10^{-4}~P_0\Omega H^3$ ($\ell_v=1.42~P_0\Omega$), while all other parameters (e.g. resolution, offset of the body from corotation) are that of the fiducial simulation L1K1. The value of $\lambda/H$ is easily read off; the value of $x_p/\lambda_c$ is obtained by noting that $x_p=0.097~H$. * indicates a simulation that has density fluctuations greater than 5\% of the equilibrium value and has thus entered the non-linear regime. Simulation L1K1 is often referred to as the fiducial simulation, and L1K10 as the high conductivity simulation. Measured values are presented as the average between 1 and 10 orbits plus/minus one standard deviation of the value in time. \label{tab:fixedLsims}}
\tablewidth{0pt}
\tablehead{
\colhead{} & \colhead{$\chi~[H^2\Omega]$} & \colhead{$\lambda_c/2~[H]$} & \colhead{$F_y~[H\Omega^2]$ (Linear Prediction)} & \colhead{$F_y~[H\Omega^2]$ (Measured)}
}
  \startdata
* & 5.97$\times10^{-3}$ & 1.54$\times10^{-1}$ & 1.36$\times10^{-2}$ & 7.96$\times10^{-3}\pm4.24\times10^{-5}$ \\
* & 6.14$\times10^{-3}$ & 1.56$\times10^{-1}$ & 1.31$\times10^{-2}$ &  7.79$\times10^{-3}\pm3.98\times10^{-5}$ \\
 & 8.53$\times10^{-3}$ & 1.84$\times10^{-1}$ & 7.97$\times10^{-3}$ &  5.95$\times10^{-3}\pm2.61\times10^{-5}$ \\
 & 1.02$\times10^{-2}$ &        2.01$\times10^{-1}$ & 6.07$\times10^{-3}$ &  5.06$\times10^{-3}\pm2.04\times10^{-5}$ \\
 & 1.19$\times10^{-2}$ &        2.17$\times10^{-1}$ & 4.81$\times10^{-3}$ &  4.37$\times10^{-3}\pm1.64\times10^{-5}$ \\
 L1K1 & 1.71$\times10^{-2}$ &        2.60$\times10^{-1}$ & 2.82$\times10^{-3}$ &  3.03$\times10^{-3}\pm 9.46\times10^{-6}$ \\ \hline
 & $1.72\times10^{-2}$ &        2.61$\times10^{-1}$ & 2.78$\times10^{-3}$ &  3.00$\times10^{-3}\pm9.46\times10^{-6}$ \\
 & $1.88\times10^{-2}$ &        2.72$\times10^{-1}$ & 2.44$\times10^{-3}$ &  2.74$\times10^{-3}\pm8.32\times10^{-6}$ \\
 & $3.41\times10^{-2}$ &        3.67$\times10^{-1}$ & 9.97$\times10^{-4}$ &  1.37$\times10^{-3}\pm3.54\times10^{-6}$ \\
 & $5.12\times10^{-2}$ &        4.50$\times10^{-1}$ & 5.43$\times10^{-4}$ &  8.24$\times10^{-4}\pm2.17\times10^{-6}$ \\
 & $8.53\times10^{-2}$ &        5.80$\times10^{-1}$ & 2.52$\times10^{-4}$ &  4.21$\times10^{-4}\pm1.35\times10^{-6}$ \\
 L1K10 & $1.71\times10^{-1}$ &        8.21$\times10^{-1}$ & 8.91$\times10^{-5}$ &  1.65$\times10^{-4}\pm8.04\times10^{-7}$ \\
 L1K50  & 8.53$\times10^{-1}$ &        1.84 & 7.97$\times10^{-6}$ &  1.93$\times10^{-5}\pm2.92\times10^{-7}$ \\
    \enddata
\end{deluxetable*}

\subsection{Numerical Considerations} \label{ssec:numerical}
In order to assess the robustness of the numerical results, we explored the dependence of the simulation results on domain size, resolution, boundary conditions, and injection radius. Of these factors, we find that the most important numerical effects are related to the size of the injection region. The analytic assumption that all the body's luminosity is deposited at a single point is both an approximation to the physical situation, and an idealization that cannot be achieved in grid-based numerical simulations. We find that for the fiducial parameters and a resolution that allows for an injection radius of $r_\mathrm{rad}=0.04~H$, the measured azimuthal force is $7.6\%$ larger than the linear theory's prediction. Doubling the injection radius to $r_\mathrm{rad}=0.07~H$, at half the resolution, leads to an error with respect to the linear prediction of $-25\%$, i.e. a decrease in resolution results in a measured azimuthal force smaller than the linear theory's prediction. An even higher resolution with a correspondingly small injection radius could result in even better agreement with the linear theory; however, at this point the question of more detailed physics close to the body would likely be more pressing.

To isolate the effect of changing spatial resolution from the effect of differing injection radii, we test for convergence with spatial resolution by keeping the same injection radius and changing the resolution. Due to the discretization of the region around the body, increasing the resolution will result in an injection region that closer approximates a sphere rather than a rectangular prism (as is the case for the low resolution simulation, which injects energy evenly into eight neighboring cells). Because of the slight change in injection volume, the total injected luminosity will also be modified; since we have an excellent prediction of what a simulation with a slightly different total luminosity would be (see Fig.~\ref{fig:fixedchi}), we can control for the difference in total luminosity and isolate the influence of the injection region's shape. We compare two simulations, both with an injection radius of $0.07~H$, conductivity $\chi=6.1\times10^{-3}~H^2\Omega$, and injected luminosity per volume $\ell_v=1.42~P_0\Omega$ but one with fiducial resolution and the other with half the fiducial resolution, resulting in total injected luminosity $2.5\times10^{-4}~P_0\Omega H^3$ and $1.5\times10^{-4}~P_0\Omega H^3$, respectively. We find that the net force per unit mass agrees between these runs at approximately the 10\% level. (Note that for this value of the conductivity neither the high nor the low resolution simulation recover the analytic prediction to high accuracy.) 

From Fig.~\ref{fig:fy-rad} it is apparent that the heating torque arises from within approximately $0.5H$ of the body for simulation L1K1, well within the size of our fiducial simulation domain. Nonetheless, the use of periodic boundary conditions does introduce artefacts that are visible in the plots of the density and pressure perturbations as structures close to the edges of the box that re-appear on the opposite side of the box (sheared, in the case of the y-edges; see Fig.~\ref{fig:lin-nonlin}). To test for possible errors introduced by the use of periodic boundaries, we compared simulations in which the box size was increased to twice that of the fiducial simulation L1K1's in each direction, while maintaining all other variables constant. The measured accelerations agreed to better than 1\% at every point in time, including the time (between 0.4 and 0.7 orbits) when density perturbations had re-entered the fiducial simulation domain but had not yet reached the edge of the doubled simulation domain, as well as the steady state at later times. The same holds when the box size of L1K10 is doubled. Somewhat larger changes, at the 5\% level, occur if we compare against a box with half the resolution, but three times the box size, of the fiducial simulation L1K1, likely due to the aforementioned increase in the injection region. The lack of impact that the artificially heated gas has on the measurement of the azimuthal acceleration is because the density perturbations near the edges of the box are on the order of two orders of magnitude lower than the regions closest to the body (Fig.~\ref{fig:lin-nonlin}). The combined effects of the low magnitude of the density perturbations and their larger distance from the body, which reduces their contribution to the net azimuthal force, suggest that the density and pressure perturbations that exit and re-enter the box through artificial periodic boundary conditions do not impact the force calculation at the level of accuracy we are interested in here. For the extremal simulation L1K50, however, the perturbations are of the same magnitude close to and far away from the body, resulting in the increased deviation from the linear prediction seen in Fig.~\ref{fig:fixedLtot}.

Finally, we note that the simulations assume that the luminous body's location remains fixed over a small multiple of the local dynamical timescale. In principle, for sufficiently high luminosities and local disk surface densities, the resulting torque might be able to migrate the body fast enough to invalidate this assumption. Analogous physics has been studied in the context of gravitational torques, where motion of the gravitating body can lead to a dynamical co-rotation torque and ``Type III" migration \citep{masset03,paardekooper14}. We do not explore this possibility here, but note that caution and additional study would be needed in any circumstance where the implied migration speed due to the heating torque exceeded a fraction of $H \Omega$.

\section{Discussion} \label{sec:discussion}
\subsection{Limits of the Linear Theory} \label{ssec:limitations}
The analytic theory for the heating torque relies both on linearity, and on satisfying hierarchical separation between the scales of the displacement from co-rotation, the induced density perturbation, and the disk scale height. By numerically solving the full set of hydrodynamic equations, we can test the limits of these various assumptions.

\subsubsection{Testing the Hierarchy Requirements} \label{sssec:hierarchyTest}
The first set of assumptions are the hierarchies given by Eq.~\ref{eq:hierarchyI} and Eq.~\ref{eq:hierarchyII}, i.e. that  $x_p\ll \lambda_c\ll H$. Fig.~\ref{fig:fixedLtot} shows how the derived acceleration scales as these assumptions are broken. The simulations closest to the analytic prediction do not have equal ratios of $x_p/\lambda_c$ and $\lambda_c/H$; rather, they prefer a smaller $x_p/\lambda_c$. As thermal conductivity increases, $\lambda_c$ becomes larger whereas the offset from corotation $x_p$ and disk scale height $H$ stay constant. This results in a decrease in the ratio $x_p/\lambda_c$ and an increase in $\lambda_c/H$, i.e. Eq.~\ref{eq:hierarchyI} becoming better satisfied and Eq.~\ref{eq:hierarchyII} becoming less satisfied. The result is that the characteristic wavelength of the perturbations is less well-contained by the simulation domain, particularly in the case of L1K50, whose characteristic wavelength of $3.6~H$ does not fit in the simulation domain at all.  Extending L1K50's simulation domain to a scale height where all three scales are separated by a factor of 10 (i.e. the offset from co-rotation is a factor of 100 smaller than the scale height) would require upwards of 1200 cells in each direction to resolve a single scale height, which is not even large enough to capture the full decay of higher conductivity simulations. The computational cost of such simulations is beyond the scope of this work. 

\begin{figure}
    \centering
    \includegraphics[width=0.48\textwidth]{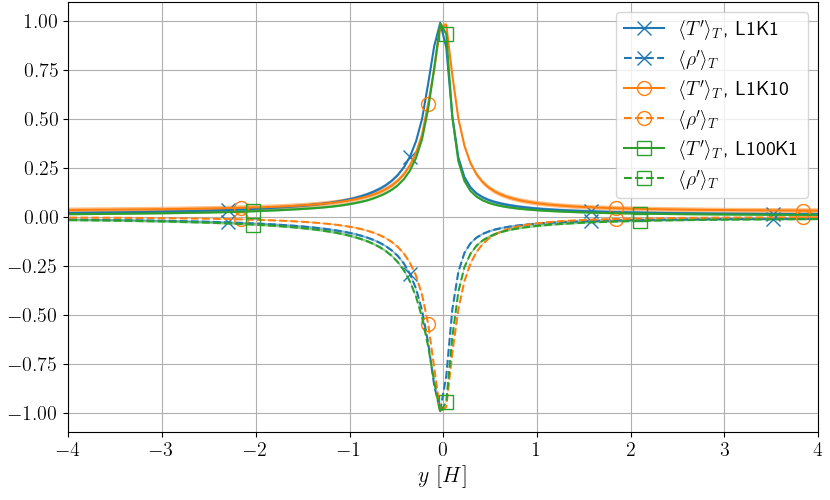}
    \caption{Time-averaged azimuthal profiles at the body's position of the density (dashed line) and temperature (solid line) perturbations $\rho'$ and $T'$ for the fiducial simulation L1K1 (blue crosses), and a high conductivity simulation L1K10 (orange dots), and a high luminosity simulation L100K1 (green squares). Each line has been normalized to its maximum value ($0.14$ and $0.15$ times L1K1's density and temperature perturbation maxima for L1K10; $81.3$ and $137.6$ times L1K1's density and temperature perturbation maxima for L100K1) and time-averaged over the last seven orbits, $t=3$ to $9.6$ orbits. Filled-in portions denote one standard deviation over time.}
    \label{fig:denTcorr}
\end{figure}%

\subsubsection{Non-linear Effects}\label{sssec:non-linear}
Another assumption is that the perturbations are small compared to their equilibrium values: $\rho'\ll\rho_0$. By increasing the luminosity, we can study how the acceleration departs from the linear prediction as we enter the non-linear regime where $\rho'\sim\rho_0$. The scaling relation of acceleration with fixed conductivity (Fig.~\ref{fig:fixedchi}) shows that higher luminosity simulations that are in the non-linear regime (indicated by green squares) measure a smaller acceleration than the linear theory would predict. Because the role of periodic boundary conditions was determined to be negligible in \S\ref{ssec:numerical}, this difference could be due to either the computational issues with resolving hierarchies listed in the previous section or to non-linear effects that are not adequately captured by the linear theory. A term is considered non-linear if it contains the product of two or more perturbations, e.g. $\rho'T'$, and is hence neglected in the linear theory, which only carries the perturbations to first order under the assumption that the perturbation is much smaller than its equilibrium value. The interaction of the linear perturbations $\rho'$ and $T'$ through terms such as the heat flux (described below) provide a mechanism through which the simulations presented (which solve the full set of hydrodynamic equations) can differ from the linearized model of \citet{masset17}. In this section we show that the measurement of the difference in azimuthal acceleration is physical, i.e. that the linear theory's neglect of higher-order terms leads to an overprediction (underprediction) of the actual non-linear net azimuthal acceleration for L100K1 (L1K10).

\begin{figure}
    \centering
    \includegraphics[width=0.48\textwidth]{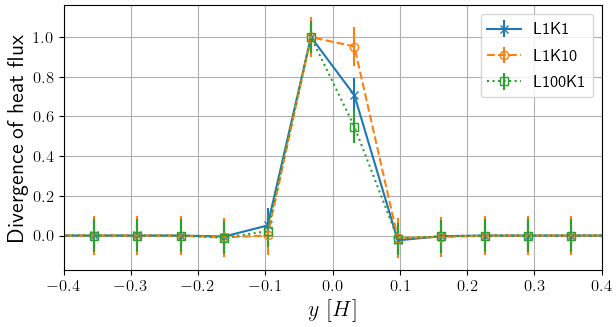}
    \caption{Time-averaged azimuthal profiles at the body's position of the divergence of the heat flux's non-linear contribution $\nabla\cdot\left[\chi\rho'\nabla T'\right]$ for the fiducial simulation L1K1 ($\ell_v=1.42~P_0\Omega$, $\chi=0.017~H^2\Omega$; blue crosses), the high conductivity simulation L1K10 ($\ell_v=1.42~P_0\Omega$, $\chi=0.17~H^2\Omega$; orange circles), and the high luminosity simulation L100K1 ($\ell_v=142~P_0\Omega$, $\chi=0.017~H^2\Omega$; green squares). Each line has been normalized to its maximum value ($0.004$ and $2.9\times10^4$ times the fiducial simulation's maximum value, respectively) and time-averaged over $t=3$ to $9.6$ orbits. Error bars denote one standard deviation over time. }
    \label{fig:heatflux}
\end{figure}

There are two properties of the steady-state perturbations that could contribute to the final density distribution: their profiles' shapes and their amplitude. Time-averaged azimuthal profiles of the density perturbation (Fig.~\ref{fig:denTcorr}) in which the high luminosity (L100K1) and fiducial (L1K1) simulations' profiles lay directly on top of each other once normalized demonstrate that the shape for these two simulations is not contributing to the measured differences. Similarly, Fig.~\ref{fig:heatflux} shows the non-linear term $\nabla\cdot\left[\chi\rho'\nabla T'\right]$ (the divergence of the heat flux's contributions by non-linear terms; this term is subtracted from the time derivative of the internal energy density in Eq.~\ref{eq:energyeq}). The normalized profiles of the heat flux from perturbed quantities of both the fiducial (L1K1) and high luminosity (L100K1) simulations overlay one another within one standard deviation. These similarities in shape show that the relevant length scales are indeed the same for these two simulations, further suggesting that it is not the shape but rather the amplitude of the profiles that contributes to the measured differences. Indeed, the magnitude of the higher luminosity run's heat flux is four orders of magnitude larger than the fiducial simulation. In contrast, the azimuthal density and heat flux profiles of the high conductivity run L1K10 (which has a larger characteristic wavelength $\lambda_c\propto\chi^{1/2}$) do not line up with the fiducial and high luminosity runs even when normalized. In the high conductivity run, the heat more readily diffuses from its injection region to the $y>0$ region, resulting in a broader high temperature region with a lower maximum temperature compared to the L1K1 and L100K1 simulations (Fig.~\ref{fig:denTcorr}). For the same reason of heat diffusion, the azimuthal profile of density perturbations in Fig.~\ref{fig:denTcorr} appears more symmetric close to the body than L1K1 and L100K1. As discussed below, the increased symmetry results in a smaller net acceleration close to the body in L1K10 than in L1K1. This change in the shape of the density's azimuthal profile is not captured by the linear theory, which predicts that the three-dimensional distribution of the density perturbation does not depend on thermal conductivity~\citep[Eq. 119]{masset17}.

\begin{figure}
    \centering
    \includegraphics[width=0.48\textwidth]{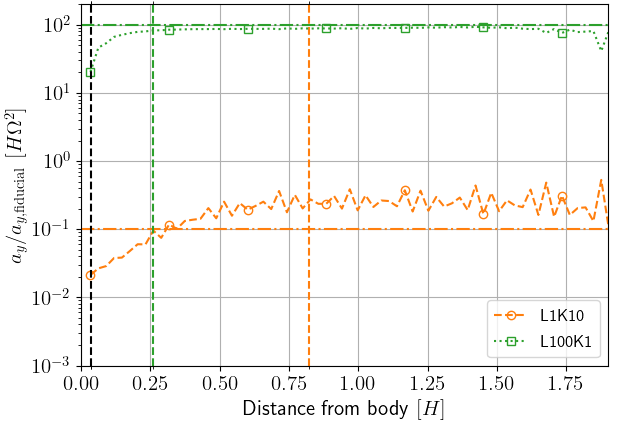}
    \caption{Snapshot of the radial profile of the net azimuthal acceleration on the body at a time of 9 orbits for a high conductivity simulation L1K10 ($\ell_v=1.42~P_0\Omega$, $\chi=0.17~H^2\Omega$; orange dash-dot line), and a high luminosity simulation L100K1 ($\ell_v=142~P_0\Omega$, $\chi=0.017~H^2\Omega$; green dotted line), normalized at every point to the fiducial simulation L1K1. As in Fig.~\ref{fig:fy-rad}, the injection radius $r_\mathrm{rad}=0.04~H$ for all three simulations (black vertical line) and half the characteristic wavelength for L1K1 and L100K1 ($\lambda_c/2=0.26~H$) and L1K10 ($\lambda_c/2=0.82~H$) are plotted. Horizontal lines are the profiles that L1K10 and L100K1 would have if they had the same shape as L1K1, i.e. the linear theory's prediction.}
    \label{fig:ygforce}
\end{figure}

 To quantify what the shape of the heat flux and density/temperature azimuthal profiles means for the acceleration of the body, Fig.~\ref{fig:ygforce} plots the net azimuthal acceleration from the gas as a function of distance from the body (very similar to Fig.~\ref{fig:fy-rad}'s plot of the sum of the azimuthal acceleration in either y-direction), normalized to the values of the fiducial run at every point. Because the linear theory predicts that only the size of the low density perturbation and not the shape should change with varying luminosity and conductivity, we use the fiducial simulation as a template to predict the azimuthal acceleration profile and scale it by $L/\chi$. These profiles are plotted as horizontal lines in Fig.~\ref{fig:ygforce}. In this figure, the high luminosity (L100K1) simulation's acceleration profile is modified only close to the body, where it is lower than might be expected from linear theory. Notably, the magnitude of the perturbations is also the highest close to the body (as seen in Fig.~\ref{fig:lin-nonlin}), suggesting that non-linear effects such as the heat flux term profiled in Fig.~\ref{fig:heatflux} cause the deviation from linear theory. On the other hand, the azimuthal acceleration on the body in L1K10 is less than the linear prediction within about half a scale height of the body as suggested by the symmetry seen in Fig.~\ref{fig:denTcorr}, but grows steadily due to higher value of the density perturbations at $y>0.5~H$ compared to those at $y<-0.5~H$. The azimuthal acceleration saturates at a value approximately twice as high as the linear theory, ultimately leading to a measured azimuthal acceleration larger than the linear prediction (see Fig.~\ref{fig:fixedLtot}). Although the increased value of the acceleration far from the body might suggest that periodic boundary conditions are artificially increasing the measured acceleration, this density distribution was achieved in larger boxes before the density perturbations re-entered the simulation domain and is thus physical (see \S\ref{ssec:numerical}), suggesting that it is indeed physics neglected by the linear theory that alter the symmetry of the density distribution.

We again note that the separation of the effect of the body's luminosity from its gravitational potential is only valid in the linear regime. In the non-linear regime, interaction between these two effects (which linearly act in the same way to provide a net outward migration) could result in deviation from the linear prediction. Exploring aspects of this interaction is left to future studies. 


\subsection{Physical Parameter Regimes} \label{ssec:scenarios}

\subsubsection{Stars or Accreting Compact Objects in a Thin Disk} \label{sssec:agn}
Migration processes may be important in geometrically thin AGN accretion disks. Stars may form within such disks as a consequence of gravitational instability \citep{goodman2003,levin07}, and they may also be captured from a cluster whose orbits intersect the disk gas \citep{syer91}. Either circumstance could lead to a population of luminous stars, or accreting stellar-mass compact objects, orbiting within an AGN disk.

The full ramifications of having a population of stellar-mass objects within AGN disks are complex, and we do not discuss them here. Rather, we assume that we have a single luminous object orbiting on a circular, non-inclined orbit, with the same sense of rotation as the disk gas. The question we seek to answer is whether the heating torque is large enough, compared to previously studied torques arising at the Lindblad and co-orbital resonances, that it should be included in models of migration within AGN disks.

We assume, consistent with our numerical results, that the analytic result given as Eq.~\ref{eq:htorquemag} provides a good estimate of the ratio of the heating torque, $\Gamma_{\rm heat}$, to the fiducial torque scaling, $\Gamma_0$. We assume a Keplerian disk ($q = 3/2$) in which the pressure $p_0 \propto r^{-n}$. At $r=r_0$ the relevant quantities can then be written as,
\begin{eqnarray}
 x_p & = & \frac{n c_s^2}{3 r_0 \Omega_0^2\gamma}, \\
 \lambda_c & = & 2 \pi \sqrt{\frac{2 \chi}{3 \gamma \Omega_0}}, \\
 L_c & = & \frac{4 \pi}{\gamma} GM \chi \rho_0.
\end{eqnarray}
Here $\chi$ is the thermal diffusivity in the disk gas surrounding the luminous object. If the diffusivity is physically the result of radiative diffusion, we can write \citep[correcting the factor of 4 typo]{paardekooper11},
\begin{equation}
    \chi = \frac{ 16 \gamma (\gamma-1) \sigma T^4 }{3 \kappa \rho_0^2 H^2 \Omega_0^2},\label{eq:paardekooper}
\end{equation}
where $\kappa$ is the opacity, $T$ is the temperature, and $\sigma$ is the Stefan-Boltzmann constant.

For both massive stars and accreting compact objects, the Eddington luminosity provides a very rough but reasonable estimate of how the likely luminosity scales with the mass. We write,
\begin{equation}
    L = \epsilon L_{\rm Edd} = \frac{4 \pi \epsilon c GM}{\kappa}, 
\end{equation}
where $\epsilon$ is an efficiency factor that may be larger than one. The above formulae then give a scaling,
\begin{equation}
    \frac{\Gamma_{\rm heat}}{\Gamma_0} \propto 
    \left( \frac{H}{r_0} \right) 
     c_s^2 \kappa^{1/2} \frac{H^2 \rho_0^2 \Omega_0^{3/2}}{T^6}.
\end{equation}
The numerical pre-factor depends upon the assumed vertical structure of the disk. Taking $\rho_0 = \Sigma / H$ and $c_s = \gamma H^2 \Omega_0^2$ we find, 
\begin{equation}
    \frac{\Gamma_{\rm heat}}{\Gamma_0} \simeq 0.053 
    \frac{n \epsilon}{\gamma (\gamma-1)^{1/2}}
    \left( \frac{H}{r_0} \right) 
    \frac{c \kappa^{1/2}}{\sigma^{3/2}} \frac{H^2 \Sigma^2 \Omega_0^{7/2}}{T^6}.
\label{eq_torque_ratio}    
\end{equation}  
There is no dependence on the mass of the luminous object. We note that 
the above analysis has assumed in various places that the disk is optically thick, that it is supported by gas pressure, and that the luminosity of the embedded object is transported out by radiative diffusion. 

Given a disk model, for example a Shakura-Sunyaev disk in one of the gas pressure dominated regimes \citep{ss73}, it is straightforward to estimate the ratio of $\Gamma_{\rm heat} / \Gamma_0$. The result is fairly complex expressions that obscure the basic question of whether $\Gamma_{\rm heat}$ can be neglected when considering migration. It is more illuminating to forego explicit reference to the opacity, and write Eq.~(\ref{eq_torque_ratio}) in a manifestly dimensionless form that involves the optical depth $\tau$. To do so we need the following results for a thin accretion disk in a steady state \citep{frank02}:
\begin{eqnarray}
    \tau & = & \Sigma \kappa, \\
    T^4 & = & \frac{3 \tau \dot{M} \Omega_0^2}{8 \pi \sigma}, \\
    \nu \Sigma & = & \frac{\dot{M}}{3 \pi}, \\
    \nu & = & \alpha c_s H.
\end{eqnarray}
(We have dropped some unimportant numerical factors from these expressions.) Using these results, and adopting reasonable values for the adiabatic index and pressure gradient parameter ($\gamma=5/3$, $n=3$) we obtain,
\begin{equation}
    \frac{\Gamma_{\rm heat}}{\Gamma_0} \sim 0.07 \left( \frac{c}{v_K} \right) \epsilon \tau^{-1} \alpha^{-3/2},
\end{equation}
where $v_K$ is the Keplerian velocity in the disk. The condition for the heating torque to be important (relative to the Lindblad and co-orbital torques, again as defined by~\cite{masset17} rather than~\cite{paardekooper11}), $\Gamma_{\rm heat} > \Gamma_0$) is then,
\begin{equation}
    v_K \tau \alpha^{3/2} \lesssim 0.07 c \epsilon. \label{eq:tauBound}
\end{equation}
In an AGN disk we expect $\alpha < 1$, and across most of the region where stars would form or be captured $v_K \ll c$. It is then clear that an embedded object, radiating a luminosity of the order of the Eddington limit ($\epsilon \sim 1$), will experience dominant heating torques at any radii where the optical depth is modest. 

As a specific example, we apply the criterion in Eq.~\ref{eq:tauBound} to the model of a constant Toomre parameter disk~\citep{goodman2003}. Taking the standard parameters presented in this model, we use a central black hole mass of $10^8$ solar masses, a radiative efficiency $\mathcal{E}=L/\dot Mc^2=0.1$, assume that the viscosity is proportional to the total pressure rather than the gas pressure, and use the electron-scattering opacity $\kappa=0.4~\mathrm{cm^2/g}$. Although this model cannot be directly mapped onto a Shakura-Sunyaev disk, it is useful as an example. From~\citet{goodman2003}'s Eq. 10, the radius at which gravitational instability sets in is at $r_{Q=1}\approx10^3 R_s$. Using a value of $\alpha=0.01$ appropriate for MRI turbulence and assuming a luminosity on the order of the Eddington limit, the resulting bound is 
\begin{equation}
    \tau\lesssim2200.
\end{equation}
Using~\citet{goodman2003}'s Eqns. 16 and 18 to calculate the temperature and surface density at $r_{Q=1}$ and the prescription in Eq.~\ref{eq:paardekooper} for thermal conductivity, the relevant length scales are 
\begin{align}
    x_p&\sim 0.02R_s\\
    \lambda_c&\sim8R_s\\
    H&\sim 10R_s.
\end{align}
The assumption that $x_p\ll \lambda_c<H$ holds for this model of a luminous object in an AGN disk, although the characteristic wavelength $\lambda_c$ and the gas scale height $H$ are the same order of magnitude, similar to the simulations presented in this work (see Table~\ref{tab:blparams}).

\subsubsection{Low-mass Planets in a Protoplanetary Disk} \label{sssec:blcompare}
Thermal torques were originally proposed and studied in the context of low-mass planets in protoplanetary disks, where under some circumstances they can be of the same order of magnitude as Lindblad torques. Prior simulations focused on this regime include the work of \citet{Lega14} and that of \citet{BL15}. There are physical differences between these simulations and those presented in this paper. In particular, in \citet{Lega14} the planet has mass but no luminosity (and hence is affected by only the cold thermal torque), while in~\citet{BL15} the planet is both luminous and massive (and hence is affected by both the cold thermal and the heating torque). Table~\ref{tab:blparams} compares parameters for these two simulations and the simulation L1K1. Although \citet{Lega14} and \citet{BL15} are more comprehensive physically than the present study, it is useful to compare the results of these two works to those presented in \S\ref{sec:results} to investigate the importance of resolving the region within one characteristic wavelength of the body. Additional studies investigate global effects such as the excitation of orbital eccentricities and inclinations~\citep{Eklund2017,fromenteau2019,chrenko17} and horseshoe streamlines~\citep{Chrenko2019}, which the local unstratified simulations in the present study do not capture; for instance the gas flow that would be significantly altered around a hot planet~\citep{Chrenko2019} is not seen in the streamlines of L100K1 plotted in Fig.~\ref{fig:lin-nonlin}. Though these global-scale simulations could be subject to the same limits of resolution as \citet{Lega14} and \citet{BL15}, they are not discussed in detail for the sake of clarity and brevity.

\begin{deluxetable*}{c|c|c|c} 
\tablecaption{Summary of important values for different simulations.~\citet{Lega14} studies the cold thermal torque (no luminosity), while~\citet{BL15} studies the cold and heating torques in a semi-global simulation. The values in the table were obtained either from Sec. 5.3.2 in~\cite{masset17} ($\dag$), by private communication ($*$; in particular $H/R=0.036$), or by calculation from parameters found in the respective work ($\S$). Unmarked values are straightforward calculations from other values. Physical values for the fiducial simulation L1K1 were calculated assuming a body orbiting at $5.2$ AU (the same as~\cite{BL15}) around a solar-mass central body. Here $\lambda_c\equiv k_c^{-1}$ rather than as in Eq.~\ref{eq:lambdac} in accordance with~\citet{masset17}'s definition.\label{tab:blparams}}
\tablewidth{0pt}
\tablehead{
\colhead{} & \colhead{\citet{Lega14}} & \colhead{\citet{BL15}} & \colhead{Fiducial Simulation (L1K1)}
}
  \startdata
$\lambda_c$ & $2~\mathrm{cells}^\dag/0.014~\mathrm{AU}^\dag$ & $2.34~\mathrm{cells}^\S/0.0238~\mathrm{AU}^\dag$ & $41.3~\mathrm{cells}/0.084~\mathrm{AU}$\\ \hline
$x_p$ & $0.85~\mathrm{cells}/0.006~\mathrm{AU}^\dag$ & $0.98~\mathrm{cells}^\S/0.01~\mathrm{AU}^*$ & $6~\mathrm{cells}/0.0122~\mathrm{AU}$\\ \hline
$H$ & $28~\mathrm{cells}/0.196~\mathrm{AU}$ & $21.3~\mathrm{cells}^\S/0.19~\mathrm{AU}^*$ & $62~\mathrm{cells}/0.126~\mathrm{AU}$ \\ \hline
$L$ [erg/s] & N/A & $6.0\times10^{27\dag}$ & $6.0\times10^{30}$ \\ \hline
$\chi~[\mathrm{cm^2/s}]$ & $1.5\times10^{15}~\mathrm{cm^2/s}^\dag$ & $4.35\times10^{15\dag}$ & $1.0\times10^{17}$ \\\hline
$\lambda_c/x_p$ & 2.33 & 2.38 & 6.88 \\ \hline
$H/\lambda_c$ & $14^\dag$ & 8 & 1.5               
    \enddata
\end{deluxetable*}

Although the results of both \citet{Lega14} and \citet{BL15} highlighted the importance of thermal effects for an accurate assessment of the migration rate, there was a significant mismatch between the quantitative values obtained and the subsequent analytic theory of \citet{masset17}. Inspection of Table~\ref{tab:blparams} and the results of \S\ref{ssec:numerical} suggest that this discrepancy may well be due to the difficulties inherent in resolving the relevant scales in a global simulation. Compared to previous simulations~\citep{Lega14, BL15}, our fiducial simulation L1K1 better resolves the characteristic wavelength of the density perturbation.~\citet{Lega14} has approximately two cells spanning $\lambda_c$~\citep{masset17}, whereas~\citet{BL15} has approximately 2.3 cells to resolve $\lambda_c$. By using a local domain, our fiducial simulation L1K1, which resolves $\lambda_c$ with 41 cells, is better poised to capture the full effect of the thermal torques, since as Fig.~\ref{fig:fy-rad} and~\ref{fig:ygforce} show, most of the contribution to the azimuthal acceleration comes from material within $\lambda_c/2$ of the body. Limited resolution could be one of the reasons that~\citet{BL15} sees a net force about an order of magnitude below the predicted value~\citep{masset17}. As for physical parameters, our fiducial simulation L1K1's luminosity is three orders of magnitude larger than those presented in~\citet{BL15}, leading to a large value for the heating torque. However, the results should continue to scale down to lower values of luminosity, where the assumption that the planet does not change its distance from corotation over the course of the simulations should be more accurate. 

Our simulations also explore a different regime in terms of the scale hierarchy. The simulation of \citet{Lega14} has $H/\lambda_c=14$, whereas our simulations have $H/\lambda_c=1.5$; similarly, in~\citet{Lega14}, $\lambda_c/x_p=2.3$ and in our fiducial simulation L1K1 $\lambda_c/x_p=6.88$ (Table~\ref{tab:blparams}). We have better scale separation between $\lambda_c$ and $x_p$ at the expense of less scale separation between $\lambda_c$ and $H$. Our closer-to-linear results support our argument that the first criterion is more essential to the linear theory than the second. As discussed in \S\ref{sssec:hierarchyTest}, the restriction on scale height is not a physical requirement but rather an ease-of-computational one; therefore this work simply explores a slightly different physical parameter regime. 

We conclude that poor resolution of the characteristic wavelength contributes to the mismatch between previous simulations' measurement of the heating force and the linear theory's prediction. This conclusion is supported by the resolution study in \S\ref{ssec:numerical} wherein a decrease in resolution led to a decrease in the measured value of azimuthal acceleration to below the linear prediction, similar to how~\citet{BL15}'s measurement was an order of magnitude smaller than the predicted value~\citep{masset17}. Using a higher resolution (possible in a local setup), we obtain agreement to within 10\% with the analytic theory. The remaining discrepancies are usually small, exceed the analytic prediction, and reduce with higher resolution. Exceptions are simulations with thermal conductivity much smaller than the fiducial value. This trend suggests that a more precise value depends on the exact physics close to the body, which would be better modelled with a self-consistent luminosity prescription to capture accretion onto the body. Compared to previous global studies, in local shearing box simulations, the relevant small scales are better resolved and the net azimuthal acceleration is within 10\% of the linear theory in the linear regime. 

\section{Conclusions} \label{sec:conclusions}
In this work, we have used three-dimensional local simulations of a zero-mass test particle in an inviscid unstratified disk to test the analytic theory for the heating torque developed by \citet{masset17}. The heating torque arises from the interaction between a luminous disk-embedded body and Keplerian shear, which distorts low-density regions that were heated by the body into asymmetric lobes that exert a net gravitational force on the planet. In the regime where the resulting density perturbations are linear, we find good agreement between the results of our direct numerical simulations and the analytic theory. We surmise that prior global simulations probably lacked enough resolution of the energy injection region, leading to an under-estimate of the magnitude of the heating torque. Going beyond the linear theory, we explored regimes of high thermal conductivity and high luminosity. We find that at high luminosity the derived torque is smaller than the linear prediction, and attribute this as being due to non-linear terms in the heat flux. In the high conductivity regime we infer a higher acceleration than predicted by the linear theory. We find that both the non-linear terms in the heat flux and computational limitations contribute to this larger value.

At the linear level the heating torque can be considered separately from other contributions to the migration of disk-embedded bodies. Although numerically convenient, there are few if any physical circumstances where gravitational \citep{kley12} and other thermal torques \citep{Lega14} would not also need to be considered. In most cases, study of these torques requires a combination of analytic theory, local, and global numerical simulations, whose results can partially be encapsulated in relatively simple torque formulae \citep{paardekooper11,jiminez17}. In the context of low-mass planet migration, using such formulae, the heating torque is estimated to be most important (relative to the sum of all other torques) for masses on the order of one Earth mass~\citep{masset17}. This study neglects additional physics to better isolate the impact of the heating torque on low-mass planet migration; inclusion of the body's gravitational potential, disk stratification, and viscosity among other physics can be included in later studies.

A second environment where the heating torque might be important is for the migration of luminous objects (massive stars or accreting compact objects with luminosities of the order of the Eddington limit) in AGN disks. Using simple scaling arguments, and the analytic theory of \citet{masset17}, we find that the heating torque is expected to provide a dominant contribution to the total migration torque at disk radii where the optical depth drops below a critical value. The heating torque may therefore impact models for the migration, trapping, and growth of objects embedded within AGN disks, and should be considered in future analyses of such systems. In the case where the disk-embedded body is itself an accreting compact object, the mechanical luminosity of outflows may also modify the local density distribution and generate a migration torque \citep{xinyu2019}.

Using local simulations on a uniform grid, we have been able to verify the analytic predictions for the strength of the heating torque at approximately (in the most favorable cases) the 10\% level. More precise tests would be possible using static mesh refinement methods, which would also allow a fuller mapping of how the thermal torque scales with the control parameters in regimes where the assumptions of the analytic theory fail. It would also be valuable to relax the assumptions of a constant energy injection rate and conductive energy transport. Simulations that consistently resolve accretion onto disk-embedded objects, and the radiative feedback that accretion produces, are challenging but are becoming increasingly feasible.

\section{Acknowledgements}
We thank Pablo Ben\'itez-Llambay and the anonymous referee for useful discussions and feedback.
AMH acknowledges support from the National Science Foundation Graduate Research Fellowship Program under Grant No. DGE 1650115. Any opinions, findings, and conclusions or recommendations expressed in this material are those of the author(s) and do not necessarily reflect the views of the National Science Foundation. PJA acknowledges support from NASA TCAN award 80NSSC19K0639. This research was carried out in part during the 2019 Summer School at the Center for Computational Astrophysics, Flatiron Institute. The Flatiron Institute is supported by the Simons Foundation. The numerical simulations and analysis were performed on the Rusty cluster of the Flatiron Institute. 

\medskip

\bibliographystyle{aasjournal}

\end{CJK*}
\end{document}